# A Hydrogenated Amorphous Silicon Photodiode indirect flexible device for radiation flux measurements at low intensities.

Mauro Menichelli[a], Aishah Bashiri[b,c], Marco Bizzarri[d,a], Lucio Calcagnile[e,f], Mirco Caprai[a], Domenico Caputo[g,h], Anna P. Caricato[e,f], Sara M. Carturan[i,j], Massimo Cazzanelli[k], Andrea Cesarini[n] , Marco Cinausero[i], Federico Cittadini[a,j], Tommaso Croci[a,j], Sylvain Dunand[l], Michele Fabi[m,n], Giorgia Franchin[i,j], Benedetta Gianfelici[a,d], Antonella Giuri[e,f], Catia Grimani[m,n], Hamza Hasnaoui[k], Giacomo Insero[o,n], Maria Ionica[a], Keida Kanxheri[a], Nicola Lovecchio[g,h], Gianluigi Maggioni[i,j], Giuseppe Maruccio[e,f], Anna G. Monteduro[e,f], Arianna Morozzi[a], Augusto Nascetti[p,h], Stefania Pallotta[o,n], Andrea Papi[a], Daniele Passeri[q,a], Maddalena Pedio[r,a], Marco Petasecca[b], Pisana Placidi[q,a], Matteo Polo[k], Marta Ponzecchi[o,n], Alberto Quaranta[k], Gianluca Quarta[e,f], Walter Raniero[i], Silvia Rizzato[e,f], Aurora Rizzo[e,f], Federico Sabbatini[n], Rabia Sandhu[b], Leonello Servoli[a], Cinzia Talamonti[o,n], Alberto Verdini[r,a], Mattia Villani[m,n], Nicolas Wyrsch[l].

*a) INFN, Sez. di Perugia, via Pascoli s.n.c. 06123 Perugia (ITALY)*
*b) Centre for Medical Radiation Physics, University of Wollongong, Northfields Ave Wollongong NSW 2522, (AUSTRALIA).*
*c) Najran University, Faculty of Arts and Sciences, Department of Physics, King Abdulaziz Rd,1988 Najran. (SAUDI ARABIA)*
*d) Dip. di Fisica e Geologia dell'Università degli Studi di Perugia, via Pascoli s.n.c. 06123 Perugia (ITALY)*
*e) Department of Mathematics and Physics "Ennio de Giorgi" University of Salento, via per Arnesano, 73100 Lecce (ITALY).*
*f) INFN Sezione di Lecce via per Arnesano, 73100 Lecce (ITALY.)*
*g) Dipartimento Ingegneria dell'Informazione, Elettronica e Telecomunicazioni, dell'Università degli studi di Roma via Eudossiana, 18 00184 Roma (ITALY).*
*h) INFN Sezione di Roma 1, Piazzale Aldo Moro 2, Roma (ITALY)*
*i) INFN-LNL Viale dell'Università 2, 35020 Legnaro (PD) (ITALY)*
*j) Dipartimento di Fisica e Astronomia Università di Padova, via Marzolo 8, 35131 Padova (ITALY)*
*k) INFN TIFPA and Dipartimento di ingegneria industriale dell'Università di Trento, Via Sommarive 9, 38123 Povo, Trento (ITALY).*
*l) Ecole Polytechnique Fédérale de Lausanne (EPFL),Photovoltaics and Thin-Film Electronics Laboratory (PV-Lab), Rue de la Maladière 71b, 2000 Neuchâtel, (SWITZERLAND).*
*m) DiSPeA, Università di Urbino Carlo Bo, 61029 Urbino (PU) (ITALY)*
*n) INFN Sezione di Firenze, Via Sansone 1, 50019 Sesto Fiorentino, (FI),( ITALY).*
*o) Dipartimento di Scienze Biomediche sperimentali e Cliniche "Mario Serio", University of Florence Viale Morgagni 50, 50134 Firenze (ITALY).*
*p) Scuola di Ingegneria Aerospaziale Università degli studi di Roma. Via Salaria 851/881, 00138 Roma. (ITALY)*
*q) Dipartimento. di Ingegneria dell'Università degli studi di Perugia, via G.Duranti 06125 Perugia (ITALY).*
*r) CNR Istituto Officina dei Materiali (IOM), via Pascoli s.n.c. 06123 Perugia (ITALY)*

## Abstract

The objective of the Photo-HASPIDE experiment is the construction and test of an indirect a-Si:H (Hydrogenated Amorphous Silicon) photo-detector plus scintillator device on a flexible substrate for the detection and measurement of particles fluxes (X-rays, electrons and protons) and for dosimetric measurements. The idea behind this experimental project lies in the utilization of Hydrogenated Amorphous Silicon (a-Si:H) as photodiode material; owing to its notable attributes of radiation hardness, light detection capability and mechanical flexibility. After the implementation of the HASPIDE experiment, which explored direct radiation detection using a-Si:H devices on a polyimide (PI) substrate, we aim to delve into indirect detection by employing these devices in conjunction with flexible and rad-hard scintillators like polysiloxane. The indirect detector design holds promise for improved responsiveness to low radiation fluxes compared to direct detection methods. The indirect a-Si:H detector should be composed of arrays of small (about 5 x 5 $mm^2$) scintillator crystals read by a-Si:H photodiodes. Through optimization of the scintillator and detector thicknesses, we expect to achieve a better performance for low minimum detectable fluxes compared to direct detection methodologies. This new detector will find application in in-vivo dosimetry during radiotherapy or hadron-therapy and also, due to its expected fast response, in FLASH therapy. Another important application will be also in Solar Physics using these devices to measure particle fluxes in solar energetic particle events.

## 1.Introduction

Hydrogenated Amorphous Silicon (a-Si:H) is a disordered semiconductor material investigated for the first time in 1969 by Chittik, Alexander and Sterling [1]. Due to the disordered structure of amorphous silicon (a-Si), not all Si–Si bonds can be saturated and therefore some unterminated dangling bonds (DBs) are present in this disordered material (at the level of $10^{19}$ bonds/$cm^3$); these DBs behave as defects and recombination centers. The inclusion of Hydrogen in a-Si generates, as a consequence, the passivation of those DBs, reducing their density down to $10^{15}$ bonds/$cm^3$. The typical hydrogen content of a-Si:H deposited by PECVD (Plasma Enhanced Chemical Vapor Deposition), which is the most used process to produce device quality material, is in the order of 10% atomic. Notably in PECVD a-Si:H material is generally deposited at temperatures around 200-250 °C and due to this relatively low temperature, the deposition of a-Si:H layers is allowed on a wide variety of substrates.

Since these devices can be used in high radiation environment, they should have an adequate radiation hardness. Radiation hardness of a-Si:H has been tested with protons up to the level of $10^{16}$ neq (1 MeV) /$cm^2$ [2], neutrons up to 5 x $10^{16}$neq(1 MeV)/$cm^2$ [3], and with X-ray photons up to 700 kGy [4] with very good results.

The HASPIDE project (2022-2025) has successfully developed and tested a-Si:H devices deposited on polyimide (PI) for direct detection of radiation. These devices have been tested for the detection and flux measurements of proton, electron and X-ray beams. Although these detectors have excellent response at high fluxes [5], they may have signal-to-noise problems at low fluxes (namely below 50 μGy/s).

In order to overcome this limit, we are developing indirect a-Si:H detectors where an a-Si:H photodiode is coupled to a plastic scintillator.

The necessity to lower the minimum detectable flux is mostly important for space weather science applications. In particular, for solar flare, gamma-ray burst (GRB), magnetar giant flare and solar energetic particle monitoring. The detection of these events is limited by direct a-Si:H sensors to large class M and X solar flares and extreme proton fluences larger than $10^8$ particles/$cm^2$ [6,7]. These indirect sensors can be successfully employed also for medical applications, because: a) the fast response of scintillators added to the faster charge collection of a thinner device can be exploited in FLASH radiotherapy and wherever a fast response is appreciated, b) a-Si:H allows for the fabrication of extremely thin photodiodes that coupled to a scintillator constitutes an ideal solution for angular independent dosimetry c) plastic scintillators can be doped with perovskite scintillating nanocrystals to create a truly tissue equivalent scintillator (doped so as to flatten the energy mass absorption coefficient ratio with respect to water).

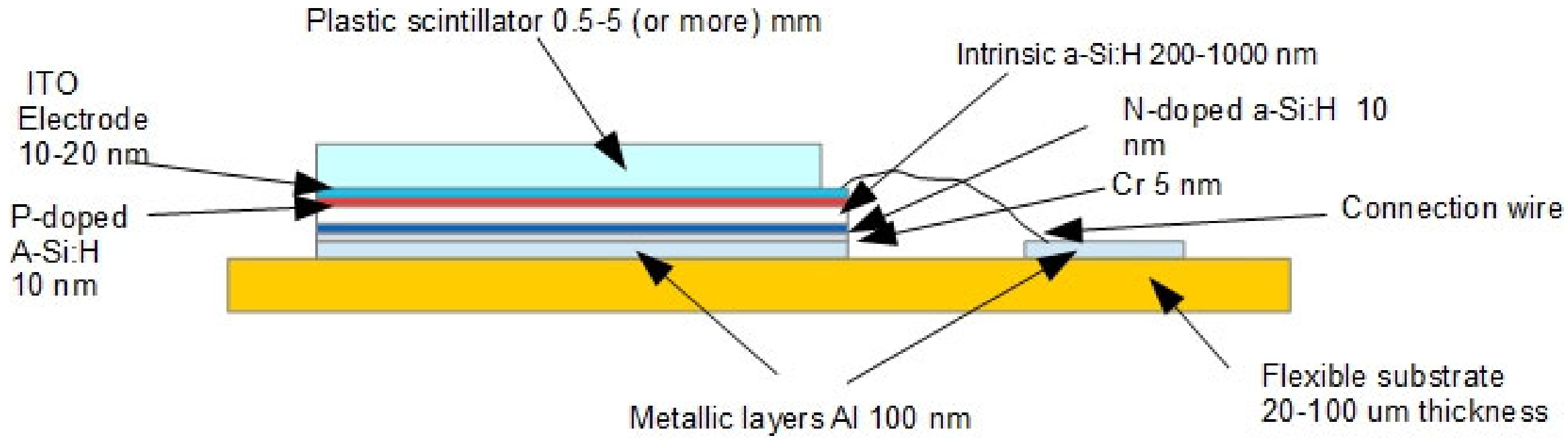

Fig. 1 Single indirect n-i-p device layout.

Figure 1 shows the basic structure of a n-i-p diode-based indirect device. The arrays of detectors will be deposited on a flexible substrate, namely polyimide, and the entire assembly should also be flexible in order to adapt to curved surfaces. The requirement of flexibility is very important for medical application in in-vivo dosimetry where this detector can be applied directly on the body of a patient. A flexibility test for the direct device that was similar to the substrate + photodetector of the photo-HASPIDE device is described in ref. [8], and shows that the device is bendable with a bending radius of about 1 cm without performance losses. Examples of previous a-Si:H indirect radiation detector could be found in refs [9,10,11] mostly for X-ray detection in medical applications. Moreover, we also aim at the optimization of the detector for charged particles radiation detection in order to allow at flux measurements for space weather gamma-ray bursts and magnetar flare detection [6,7].

As for the flexible scintillator on top of the photosensor, its role is to collect the energy deposited by impinging ionizing radiation and to convert it into emitted light whose wavelength should match as much as possible the range of sensor maximum sensitivity. Fig.2 shows the spectral response of a a-Si:H photodiode.

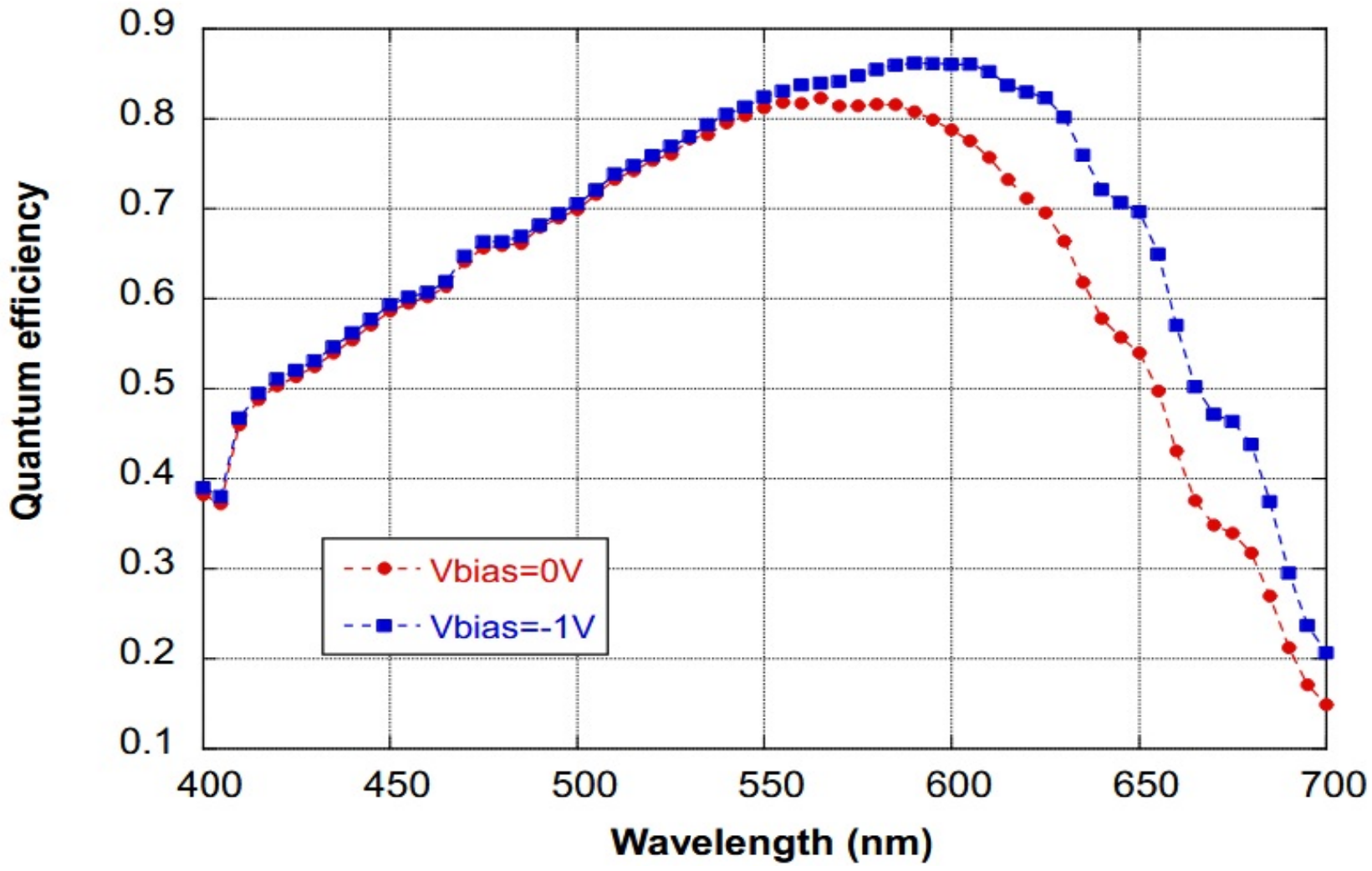

Fig.2 Quantum efficiency of a a-Si:H photodiode biased at 1 V and 0 V. It can be noted that below 550 nm there is no difference between biased and unbiased configurations while above 550 nm the quantum efficiency increases up to its maximum at 600 nm where it reaches 87% while the unbiased configuration is quite flat between 550 and 580 nm at the value of 82%.

To this aim, the main features that have to be pursued in the material design are good light yield upon radiation exposure, flexibility, radiation hardness up to 20-50 kGy, emission in the range 450-650 nm. Keeping in mind these requirements, a good choice is represented by polysiloxanes, whose features in the field of organic scintillators conveniently match the demands. As a matter of fact, polysiloxanes are intrinsically flexible as they belong to the class of elastomers. On the other hand, work is in progress within the SHINE experiment [12] to synthesize composites made of polysiloxane as flexible scaffold and inorganic Perovskites as scintillator element. Furthermore there is an ongoing activity in the Wollongong group of the photo-HASPIDE collaboration to make layered structures of scintillator + hybrid perovskite in order to optimize scintillation efficiency and tissue equivalency.

## 2. Preliminary tests on prototypes

Preliminary measurements of leakage current distribution and current versus X-ray dose rate (using a 4 W tube having a tungsten anode from Amptek at 40 kV bias voltage) has been performed on 4 a-Si:H hybrid devices glued to 4, 5 mm thick polysiloxane scintillators (similar to the detector shown in Fig.1 having 5 mm x 7 mm lateral dimensions). Three of these detectors were assembled on a device array sample (detector 4A) while the remaining one (detector 4B) was assembled on a different device array. The experimental setup (with detector 4B) is shown in Fig.3, in Fig.3a the detector 4B is placed at 13 cm from the tube collimator; the detector is glued to a PI PCB and bonded to the PCB pads, in this figure also the tube is visible. In Fig. 3b the detector is shown from the front, as it is shown there are several device pads in the detector but only one is equipped with the scintillator converter. The entire setup is hosted in a climatic chamber for thermal stabilization.

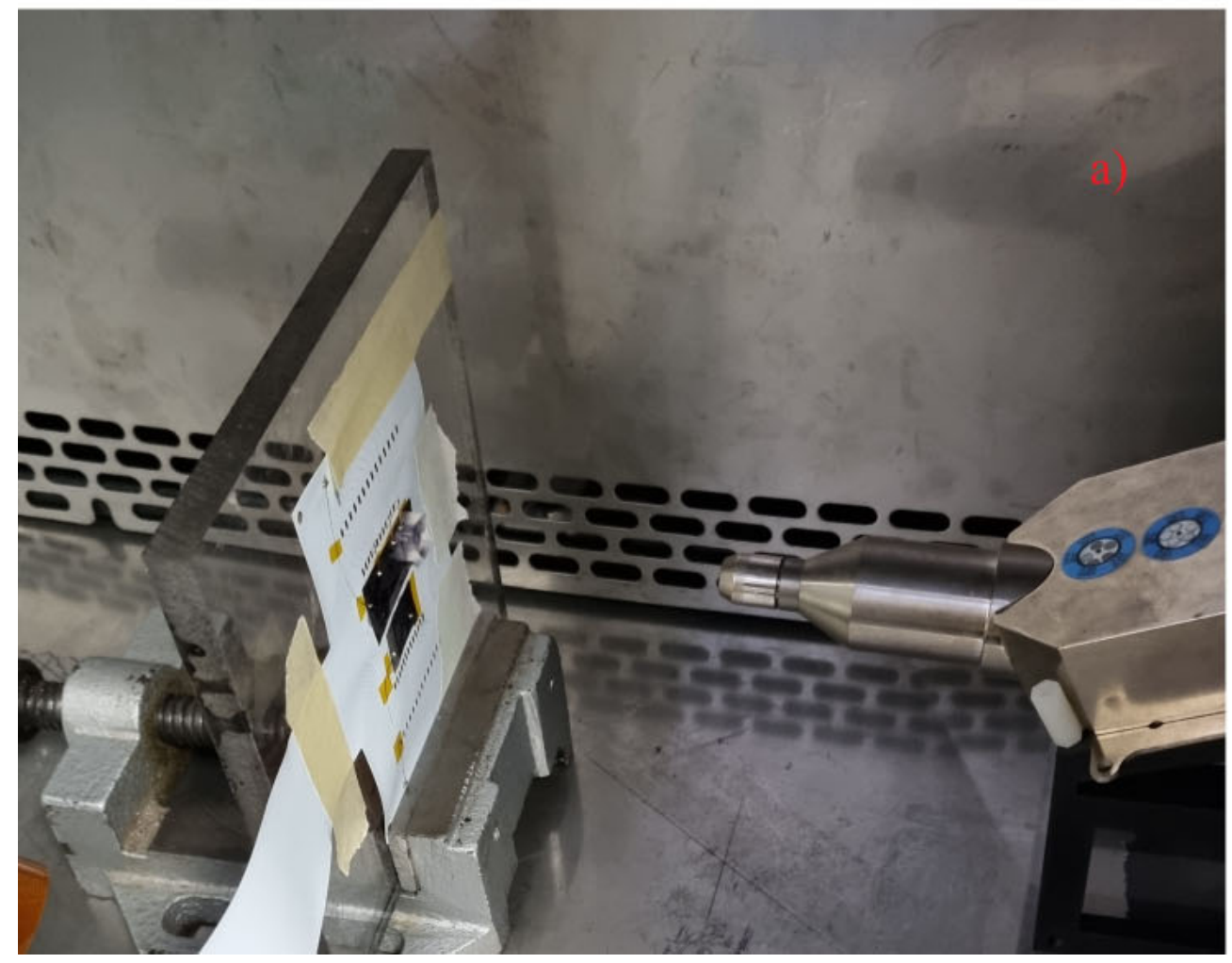

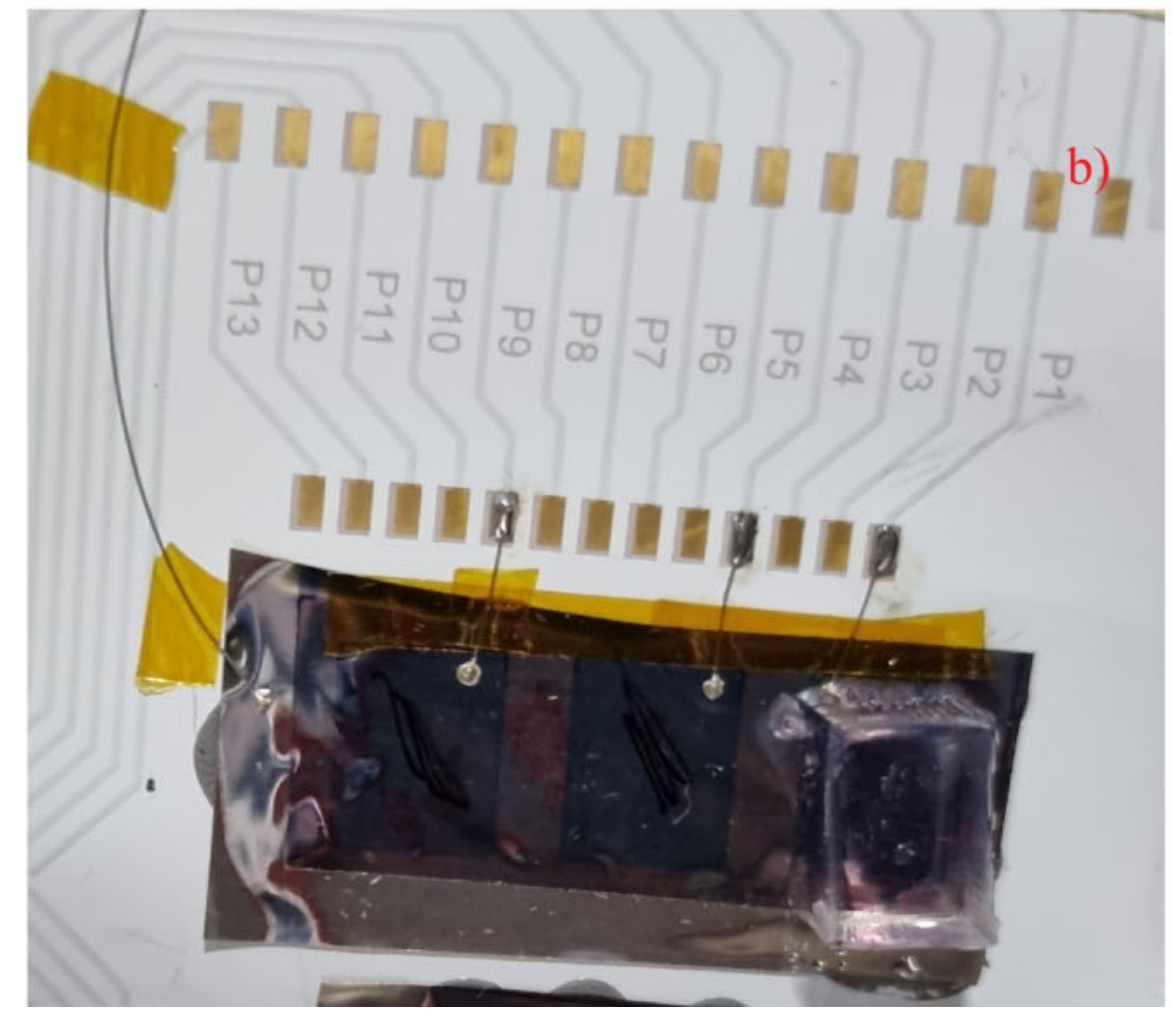

Fig.3 a) Irradiation setup with the tube shown on the right and the detector shown on the left on a test stand. b) Front view of the detector, as it is shown, there are several device pads in the detector but only one is equipped with the scintillator converter.

A preliminary calibration is performed to establish a precise correspondence between tube current and dose rate at fixed tube bias voltage (40kV), this calibration is described in [5]. Before and after the gluing of the scintillators, a measurement of leakage current versus device bias voltage (Fig.4) was performed on the 4 devices.

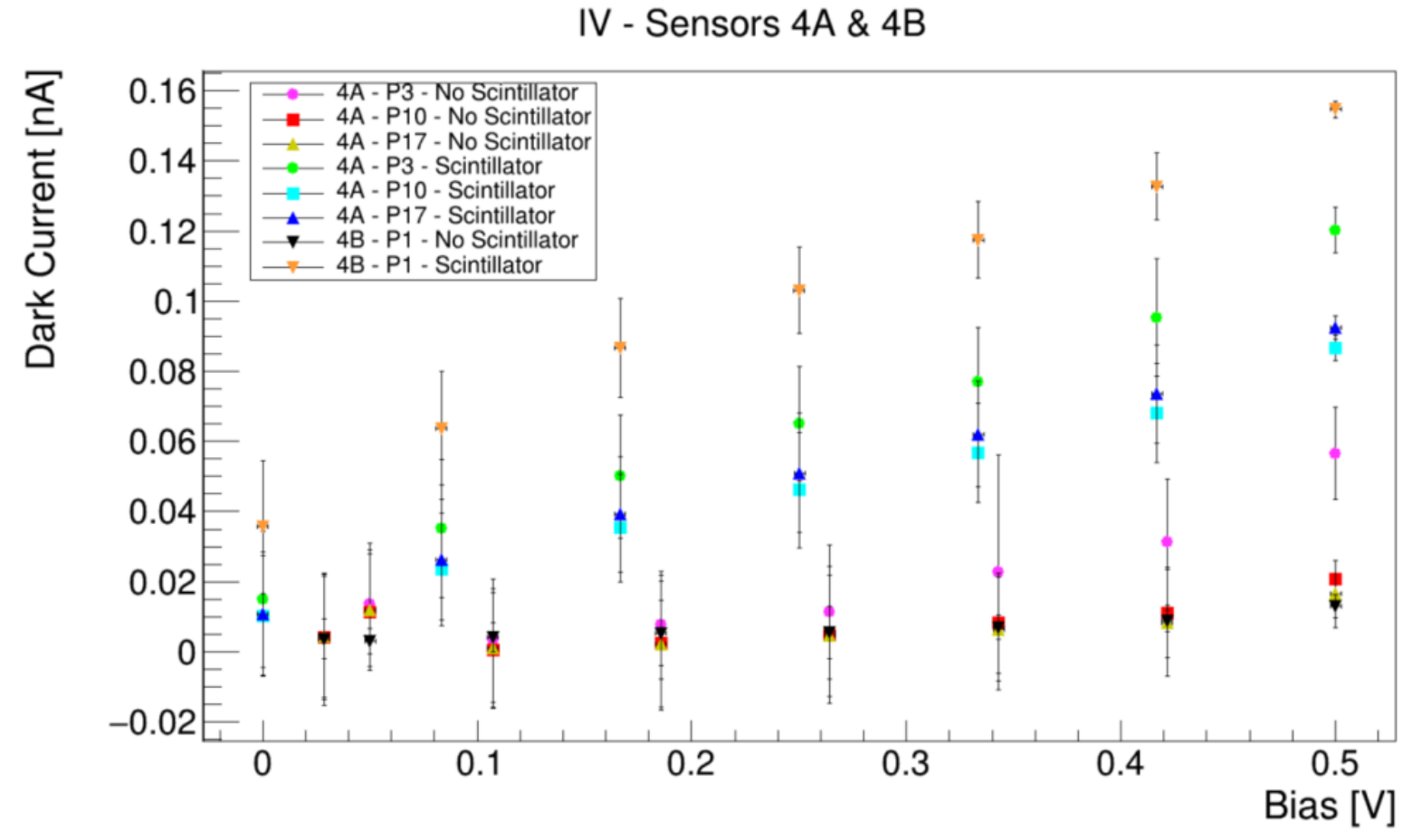

Fig.4 Dark current versus detector bias voltage with and without scintillator on the same device. At 0.5 V bias the dark current with the scintillator is larger than the bare a-Si:H device.

From Fig. 4 it is visible a notable increment of leakage current after the gluing process especially on detector 4B-P1. Fig. 5 shows the stabilization process of the leakage current in the 4 detectors tested, before and after the gluing of the scintillator when the bias voltage was set to -0.5 V. Aside an initial transient on most of the sensors, and although the value of the leakage current increases after the mounting of the scintillators the stability of the dark current tend to increase.

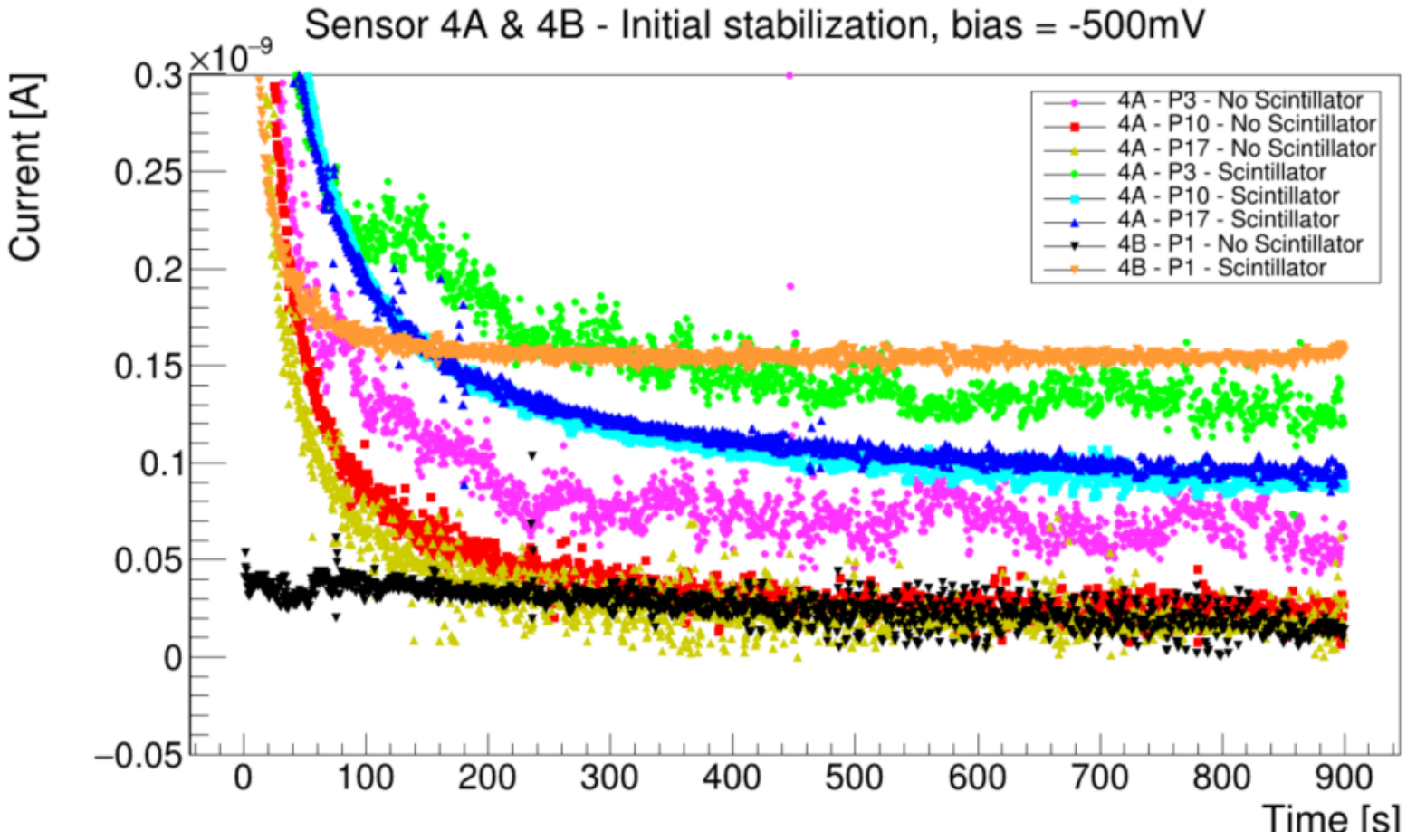

Fig.5 Stabilization of leakage current in the detector with no scintillator and with scintillator, although the leakage current is larger after the gluing, aside the initial transient, its value is more stable.

The current versus dose rate (after the subtraction of leakage current) with and without the scintillator is shown in Fig. 6 where a good linearity is shown in both configurations and, as expected, the device with scintillators shows a larger sensitivity (i.e. the slope of the line), as shown in Table I.

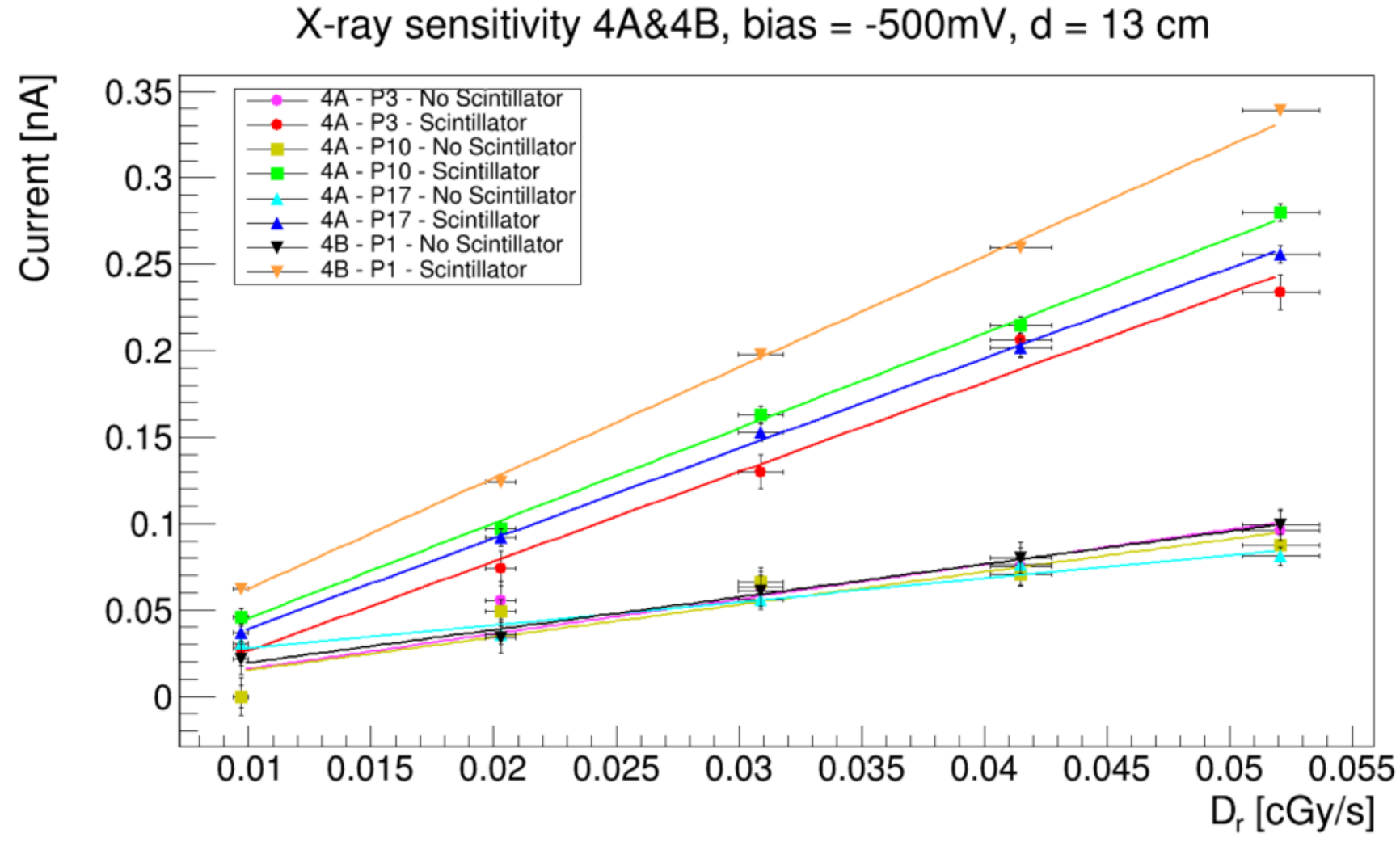

Fig.6 Output current versus dose rate for the detector biased at 0.5 V before and after the gluing of the scintillator. The behavior of the detector is linear before and after the gluing but after the gluing radiation sensitivity is larger .

| **Detector name** | **Sensitivity without Scint. (nC/cGy)** | **Sensitivity with Scint. (nC/cGy)** | **Sensitivity ratio With/without Scint.** |
|---|---|---|---|
| 4A-P3 | 2.01 ± 0.34 | 5.17 ± 0.34 | 2.57 |
| 4A-P10 | 1.89 ± 0.21 | 5.49 ± 0.21 | 2,90 |
| 4A-P17 | 1.35 ± 0.17 | 5.21 ± 0.22 | 3.86 |
| 4B-P1 | 1.90 ± 0.28 | 6.41± 0.19 | 3,37 |

Table I. Sensitivity of the various detectors measured with/without scintillator and their ratio.

These results were obtained with a scintillator type where the matching between the emission spectrum of the scintillator was not yet optimized with the quantum efficiency performance of the photodiode shown in Fig.2.

## 3. Perspectives for the Photo-HASPIDE experiment

The HASPIDE project aimed at demonstrating that hydrogenated amorphous silicon could be a suitable sensitive material to detect and monitor different type of ionizing radiation beams. This goal has been reached both for X rays tubes, synchrotron radiation, clinical electron, photon and proton beams. Some of the results obtained during the development of the HASPIDE project that will be used in the Photo-HASPIDE experiment are:

1) development and optimization of PECVD process to produce a-Si:H devices, including three types of contact, n-i-p, Charge Selective Contacts and hybrid [13];

2) design of PI substrates for the device deposition, to allow for flexible and tissue-equivalent devices;

3) interfaces between devices and readout DAQ and study of the performance of the different DAQ systems when coupled to different a-Si:H devices;

4) development of a readout chip in 28 nm technology based on the current-to-frequency conversion implemented in the past for the TERAx series chips that brought to the design of the Cleopatra chips used in HASPIDE;

5) study of the leakage current and of its fluctuation (device noise) as a function of applied bias, attached DAQ, device area and volume, to define the noise levels and hence the minimum detectable signal;

6) measurement of linearity of the response and hence sensitivity as a function of bias, radiation flux, dose rate for dosimetric applications;

7) demonstration of the radiation resistance of the devices, up to a fluence of $10^{14}$ $n_{eq}/cm^2$ with protons and of the effect of annealing procedures in recovering the performance of devices before irradiation;

Since the a-Si:H n-i-p diode has already been tested as photodiode in other experiments [14] no main critical issues are foreseen for this experiment. The main difference between the two developments (HASPIDE and Photo-HASPIDE) is that the a-Si:H device in the case of photo-HASPIDE is used as light detector from a scintillation converter, therefore the device will be optimized in a different way (thin detectors < 1 μm instead of thick detectors > 2.5 μm, transparency of the upper surface in contact with the scintillator). The application field in terms of flux measurement, in the case of Photo-HASPIDE is more optimized for low radiation fluxes detection instead of the relatively large radiation fluxes of HASPIDE.

In order to design, produce, qualify and test for applications the detectors described above the following activities are foreseen:

- Design, production and qualification of the a-Si:H photosensors.
- Design, production and qualification of the scintillators and integration with the photosensor.
- Qualification and radiation test of the assembled detector. Simulation of the detector and photosensor response.
- Medical physics application-oriented device optimization; characterization in specific clinical scenarios.
- Space weather and astrophysics application-oriented tests.

The fabrication of the a-Si:H photosensors will take place in the Department of Information Engineering, Electronics and Telecommunications at the University of Rome “La Sapienza” and at the EPFL (Neuchâtel) PV-Lab. The Rome group can also make quantum efficiency measurements on devices. Scientists at the Perugia INFN, in collaboration with those at CNR (Istituto Officina dei Materiali) can make spectroscopic analysis of the samples in terms of the correlation between electronic-properties and morphology. This activity based on the state-of-the-art characterization of electron and absorption spectroscopies, has been successfully applied to real devices made by a-Si:H [15] and can be extended to the advanced new scintillators involved in the project [12].

For the production of thin layer of polysiloxane hybrid scintillators, the reaction of hydrosilylation will be exploited, thus proper mixing of hydrosilane and vinyl terminated polysiloxane, bearing phenyl substituents, will be carried out. Then, suitable fluorophores to achieve emission will be dissolved in specific amounts and energy transfer between the matrix and the dyes will be investigated by fluorescence spectroscopy. The curing process is boosted by Pt catalyst and mild thermal conditions ($T<70°C$). In order to prepare the scintillator layers (thickness 5 mm), the tape casting process will be adopted, whereas thicker films can be obtained by casting in silicone moulds; in both cases the resin can be cured in an air oven for 8-10 hours.

As for the doping of perovskite nanoparticles this will be pursued by following the protocols defined within the SHINE project and currently under development by the LNL, TIFPA and INFN-Lecce units in a co-operative schedule. University of Wollongong will develop an alternative perovskite doped scintillator R & D. Light yield measurements upon irradiation with calibrated alpha and gamma sources will be performed by coupling the polysiloxanes either in hybrid or composite form to a photomultiplier tube, thus allowing to gather the pulse height spectrum, which will be compared with standard plastic scintillator, namely EJ-212 (Eljen Technology).

Integration of the scintillator with the photosensor can be pursued by coupling the siloxane to ITO substrate or directly on a-Si:H by using a ultrathin layer of the same base polysiloxane resin deposited by spin coating as adhesive layer. Alternatively, commercial optical glue can be adopted. Adhesion tests under bending will be performed using a bending bench where different curvature radii can be applied while observation with an optical microscope can reveal possible failure in coupling.

The test on the integrated detector will be performed in Perugia, where the detector will be glued and bonded to a Polyimide PCB for connection to readout electronics. The complete sensor will be then tested electrically (dark current versus voltage measurements) and with X-rays (dosimetric sensitivity); also a bendability test is foreseen at this stage with the setup described in [6]. Radiation tests will be performed in Wollongong with high dose rate irradiation with X-rays at the Synchrotron (ANSTO), at CEDAD (Lecce) with 3 MeV protons and in Ljubljana (SLO) with neutrons.

The simulation of detector response will be performed in Perugia with the TCAD simulation Suite previously adapted in the framework of the 3D-SiAm and HASPIDE projects. Steady-state and small-signal analyses will be performed depending on the geometry and technological parameters of the devices. Optical simulation within the scintillating layer can be properly coupled to the device simulation methods that are used to compute the optical generation rate when an optical wave penetrates into the device itself. Time-varying (i.e. transient) analysis will be also performed.

Medical application tests on the detectors (both arrays and single devices) will be primarily performed in Florence and Wollongong but also in other facilities like electrons FLASH beams. The initial tests with the photo-Haspide detectors will focus on the dosimetric characterization of detectors with different characteristics. Sensitivity will be studied as a function of scintillator thickness and voltage. Clinical beams of 6, 10, and 18 MV X-rays and electrons from 4 to 15 MeV will be used. The linearity of the signal will be verified with respect to changes in dose and dose rate. Measurements for the characterization of detectors for brachy-therapy applications will be carried out, focusing on reproducibility, linearity, and correlation with the source dwell time.

Astrophysical application tests will be performed in X-ray facilities, protons and electrons accelerators. The goal for this application is to test the response of a prototype detector consisting of active and passive layers exposed to the expected radiation fluxes in a space experiment and they will be performed in various accelerators like electron facilities (Florence, LNF-Frascati and Perugia), X-ray facilities (Perugia, Florence and Wollongong) and Proton accelerators, (Lecce, LNL, Bern, CNAO and TIFPA). The INFN Florence group will also provide FLUKA and GEANT4 simulation of the performances of this instrument for the detection of solar events, gamma-ray bursts and magnetar flares.

## 4. Conclusions

In the present paper the Photo-HASPIDE experiment for the construction of flexible a-Si:H radiation sensors has been presented. A preliminary test on an early non-optimized detector prototypes shows that the increment in sensitivity of an indirect detector built from a 0.5 µm thick a-Si:H sensor + 5 mm thick scintillator has about 3 times the sensitivity of the same device used in direct mode. A further optimization of the scintillator mostly based on the tuning of the wavelength of the emitted light with the quantum efficiency of the device and, alternatively, the perovskite + scintillator combined radiation converter adoption will hopefully enhance the sensitivity of the complete device. During and after the optimization of the device application test in the field of dose monitoring in clinical accelerators especially electron FLASH machines will be performed and also tests with protons, x-ray, and electrons will be performed for space applications.

## References.